# Josephson-Coulomb drag effect between graphene and LaAlO$_3$/SrTiO$_3$ interfacial superconductor

Ran Tao[1,2#], Lin Li[1,2#]*, Hong-Yi Xie[3]*, Xiaodong Fan[1,2], Linhai Guo[1,2], Lijun Zhu[1,2], Yuedong Yan[1,2], Zhenyu Zhang[2], Changgan Zeng[1,2]*

[1]*CAS Key Laboratory of Strongly-Coupled Quantum Matter Physics, and Department of Physics, University of Science and Technology of China, Hefei, Anhui 230026, China*

[2]*International Center for Quantum Design of Functional Materials (ICQD), and Synergetic Innovation Center of Quantum Information and Quantum Physics, University of Science and Technology of China, Hefei, Anhui 230026, China*

[3]*Division of Quantum State of Matter, Beijing Academy of Quantum Information Sciences, Beijing 100193, China*

[#] These authors contributed equally.

* Correspondence and requests for materials should be addressed to C. Z. (cgzeng@ustc.edu.cn), L. L. (lilin@ustc.edu.cn) and H.-Y. X. (xiehy@baqis.ac.cn).

**Coulomb drag refers to the phenomenon that a charge current in one electronic circuit induces a responsive current in a neighboring circuit solely through Coulomb interactions[1-3]. For conventional interactions between fermionic particles such as electrons, the as-induced drag current in the passive layer is orders of magnitude weaker than the active current due to strong dielectric screening effect between the two[4-8]. Here we propose a 'super' Coulomb drag effect between an active normal conductor and a passive superconductor of Josephson junction arrays, whereby the passive current can greatly exceed the active. The drag force originates from the interactions between the substantially enhanced dynamical quantum fluctuations of the superconducting phases in the passive layer and the normal electrons in the active layer. We demonstrate this effect in the devices composed of monolayer graphene and LaAlO$_3$/SrTiO$_3$ heterointerface, an inherently non-uniform superconductor described by Josephson junction arrays[9-11]. Remarkable drag signal is observed in the superconducting transition regime of the LaAlO$_3$/SrTiO$_3$ interface, with its sign independent of the carrier type in the graphene layer. The estimated passive-to-active ratio can reach about 0.3 at the optimal gate voltage and the temperature dependence follows that of the typical Josephson energy between superconducting**

**puddles. Strikingly, the ratio ought to be as large as $10^5$ at zero temperature by theoretical extrapolation. From engineering perspective, our device may work as current or voltage transformers, and the drag mechanism lays the foundation for synchronizing Josephson-junction-array-based terahertz radiators[12,13].**

Drag experiments have been instrumental in uncovering electron many-body effects in low dimensions as diverse as frictions between isolated two-dimensional (2D) electron gases[4-6], Luttinger-liquid and Wigner-crystal states in quantum wires[7,8], and interlayer phase coherence like excitonic superfluidity[14-18], among others. The passive-to-active ratio (PAR) offers a dimensionless parameter measuring the drag effects, whose magnitude, however, is normally far below unity[4-6]. Equal-amplitude active and passive currents may be achieved only when strong intercircuit correlations occur, such as the formation of exciton-like states[14-18]. Moreover, the sign of PAR is always exploited to distinguish various phases and interaction mechanisms[3,7,8,19].

Replacing one or both of the conducting layers with superconducting (SC) material opens opportunities for examining superconductivity and fluctuation effects[20-29]. In particular, it was proposed that Coulomb interactions between two SC layers may give rise to the non-dissipative supercurrent drag effect[23], which was originally predicted to exist in $^3$He-$^4$He mixtures[30] and neutron stars[31]. This drag effect was argued to persist even if the active layer is replaced with a normal metal. Preliminary experiments based on metal-superconductor double films were conducted two decades ago[24,25], wherein rather weak and uncontrolled drag responses were observed in the immediate vicinity of the SC transition. The estimated magnitude of PAR in metal-superconductor system is down to $\sim 10^{-3}$, and the mechanism remains unclear[23-26].

Benefiting from the fast-developing 2D material science, the sandwich structures combining graphene or/and other 2D electron systems have sparked renewed interest in studying novel physics dominated by interlayer Coulomb interactions[17,18,32-34]. The highly tunable transport properties of the component layers together with the ultra-thin dielectric spacer enable us to investigate the drag effect in previously inaccessible regimes. In this work, a unique hybrid structure of graphene and LaAlO$_3$/SrTiO$_3$ (LAO/STO) heterointerface is exploited to simulate Coulomb-coupled normal conductor and superconductor in the ultimate 2D limit. Surprisingly, we observe giant and highly gate-tunable drag responses in the vicinity of the SC transition of the LAO/STO interface. We attribute this intriguing drag phenomenon to a distinguished mechanism whereby the drag arises from the effective Coulomb coupling between the quantum fluctuations of the SC phases in a superconductor and the charge densities in a normal conductor.

The schematic of the hybrid device comprising graphene and LAO/STO, denoted as G-LAO/STO, is depicted in Fig. 1a. Macro-scale graphene (~ 2 mm) grown via chemical vapor deposition was used to avoid nanofabrication-induced degradation in the electronic performance of the LAO/STO interface (Fig. 1b, see Methods for the fabrication details). The pristine 2D superconductivity of the LAO/STO interface is well maintained in the final device, manifested as the observation of a typical Berezinskii-Kosterlitz-Thouless transition[35] (see Extended Data Fig. 1 and Supplementary Information section 1 for details). Furthermore, the interfacial superconductivity of LAO/STO can be readily tuned via a back-gate voltage ($V_{BG}$) applied across the STO substrate (Fig. 1d), as similarly demonstrated in previous studies[9,36]. The impact of $V_{BG}$ on the doping level of graphene is negligible (Extended Data Fig. 2), exhibiting a perfect shielding effect of the LAO/STO layer. On the other hand, both the carrier type and density of the graphene layer can be tuned via the interlayer-gate voltage ($V_{int}$) utilizing the LAO as the dielectric layer (Fig. 1c), with negligible impact on the electronic performance of the LAO/STO interface (Extended Data Fig. 3).

The resistance-to-temperature ($R$-$T$) curves of the LAO/STO interface exhibit two-step transitions between the high-temperature normal-metal phase and the low-temperature SC phase (Fig. 1d). This feature can be resolved from the second derivative of the $R$-$T$ curves (Fig. 1e), where we assign two characteristic temperatures, i.e., $T_P$ and $T_F$, to the peaks. The two-step SC transition has been reported in previous studies and understood in the context of electronic phase separation[9,35]. In fact, in addition to the transport characterizations, direct imaging measurements have demonstrated that the 2D superconductivity at the LAO/STO interface is spatially inhomogeneous, manifested as the formation of quenched-disorder-induced SC puddles, with a typical length scale of micrometer[37]. The SC LAO/STO interface can be interpreted as a 2D Josephson junction (JJ) array[9-11], where the inter-puddle Josephson coupling plays an essential role in the formation of global phase coherence.

Within this scenario, the first resistance drop at $T_P$ can be explained as the onset of the Cooper pairing of localized electrons and thus the formation of SC puddles, and the second drop at $T_F$ corresponds to the onset of inter-puddle phase coherence impaired by thermal or quantum fluctuations. As shown in Fig. 1f, four phases of the LAO/STO interface are distinguished in the temperature-gating phase diagram: the normal 2D electron gas state, the local SC puddle state, the phase fluctuating state, and the global phase coherent 2D SC state[9]. The three critical temperatures, $T_P$, $T_F$, and the SC transition temperature $T_C$, are non-monotonic functions of $V_{BG}$, and their maxima occur at nearly identical $V_{BG}$, which corresponds to the optimal doping.

In drag measurements we applied an active current ($I_{drive}$) to the graphene layer, and measured the passive voltage drop ($V_{drag}$) at the LAO/STO interface in open circuits[3,32,33] (see Extended Data Figs. 4-7 and Supplementary Information section 2 for validity check). As schematically shown in Fig. 2a, the moving carriers in the graphene layer could result in a passive current ($I_{drag}$) at the LAO/STO interface via interlayer interactions, and net charge accumulation at two ends induces a voltage rectifying the passive current. In Fig. 2b we show the linear dependence of $V_{drag}$ and $I_{drive}$ and depict the so-obtained drag resistance ($R_{drag} = V_{drag}/I_{drive}$) as a function of $T$ for Device #1, in which the LAO layer is 5-unit-cell thick (~ 2 nm), together with the $R$-$T$ curve of the LAO/STO interface for comparison purposes. It is manifest that a negative drag resistance peak develops at $T$ ~ 195 mK, accompanying the SC transition of the LAO/STO interface, while there is no detectable drag signal when the LAO/STO interface is either in the normal state or in the SC state.

We note that the negative sign of the drag resistance implies parallel $I_{drag}$ and $I_{drive}$. These features demonstrated in Fig. 2b can be reproduced in other G-LAO/STO devices (Extended Data Fig. 8). Applying a moderate in-plane magnetic field ($B$) of 1 T significantly depresses the superconductivity of the LAO/STO interface, indicated by the broadening of the SC transition region and the decrease of $T_C$. Accordingly, the drag resistance peak broadens and shifts to a lower temperature of $T$ ~ 160 mK (Fig. 2c). As the magnetic field is further increased to $B$ = 3 T, the drag signal disappears along with concomitant quenching of the LAO/STO interfacial superconductivity (see Extended Data Fig. 9 for the results under perpendicular fields). The correlation between the drag response and the SC transition of the LAO/STO interface is further evidenced by the field dependence of $R_{drag}$, as shown in Fig. 2d, where detectable drag occurs only within the SC transition region.

The high tunability of LAO/STO interfacial superconductivity enables us to investigate such SC-related drag effect in a wider parameter space. Figure 3a shows the phase diagram of $R_{LAO/STO}$ obtained via sweeping $V_{BG}$ at different $T$. A SC dome with a maximum of $T_C \approx 180$ mK at $V_{BG}$ ~ 27 V is clearly seen, which is in agreement with previous studies[9,36]. Figure 3b shows the phase diagram of $R_{drag}$ as a function of $V_{BG}$ and $T$. Clearly, the regions where $R_{drag}$ is finite also constitute a dome-like shape surrounding the SC dome of the LAO/STO interface. The maximum $R_{drag}$ could reach up to ~ 2 Ω, which is much larger than those observed in the systems consisting of normal metal and conventional SC films[24,25]. The inherent correlation between drag signals and LAO/STO interfacial superconductivity can also be seen via comparing the $B$-$V_{BG}$ phase diagrams of $R_{LAO/STO}$ and of $R_{drag}$ shown in Fig. 3d and Fig. 3e, respectively.

More importantly, when the extracted curves of $T_F$ and $T_C$ with respect to $V_{BG}$ are overlaid with the phase diagram of $R_{drag}$ in Fig. 3b, we find that the drag signal can be detected only when $T_C < T < T_F$ in the whole range of $V_{BG}$. One can phenomenologically describe this non-monotonic temperature dependence of $R_{drag}$ as follows. In the normal region $T > T_P$, the undetectable drag signal is expected because the interlayer inelastic Coulomb scatterings between normal electrons are strongly suppressed at such low temperatures[3,4,32]. In the SC puddle region $T_F < T < T_P$, the presence of incoherent Cooper pairs in the LAO/STO interface does not significantly enhance the drag signal. Eventually, the drag effect is detectable only when the inter-puddle SC phase coherence establishes for $T < T_F$. Within the fully SC region $T < T_C$, the measured drag resistance is again negligible simply because there is no voltage drop across a superconductor.

To unveil the drag effect below $T_F$, we analyze the PAR, defined by the dimensionless coefficient $r = I_{drag}/I_{drive} = -R_{drag}/R_{LAO/STO}$, as a function of $V_{BG}$ and $T$ (Supplementary Information section 3). As demonstrated in Fig. 3c, the PAR is positive and exhibits a dome-like shape for $T < T_F$, similar as that of the drag resistance in Fig. 3b. However, the PAR is an increasing function of temperature, and becomes unattainable in the fully SC region of the LAO/STO interface where the absolute error approaches infinity[25]. Notably, the maximum value of PAR is $r \sim 0.3$ occurring at $V_{BG} \sim 20$ V and $T \sim 170$ mK, which is two orders of magnitude larger than that obtained in the $AlO_x$-Au/Ti heterostructures[25]. Strong and highly gate-tunable PAR signals are also clearly seen from the $V_{BG}$-$B$ diagram as plotted in Fig. 3f.

We clarify the drag mechanism between a normal conductor and a superconductor before analyzing the PAR data. The low-energy physics of our hybrid device below the SC puddle temperature can be effectively described by the model of 2D Dirac fermions coupling to a 2D JJ array via Coulomb interaction. Detailed theoretical analysis can be found in Supplementary Information section 4. The effective action of the system is composed of three parts $S = S_s + S_n + S_c$. First, $S_s$ characterizes the dynamics of the phase variables $\varphi_j(t)$ of the s-wave SC puddles centered at the in-plane coordinates $\boldsymbol{R}_j$. Here we introduce the Josephson energy between puddles, $E_J(T) = E_J(0)[1-(T/T_P)^2]\exp(-T/T_0)$, in which the factor $1-(T/T_P)^2$ arises from the superfluid density with $T_P$ being the puddle temperature[10,38], and the exponential factor exp(−T/T₀) captures the Cooper-pair dephasing effect with $T_0$ proportional to inverse inter-puddle distance[39]. Second, $S_n$ describes 2D Dirac fermions with the Fermi velocity $v_F$ and the charge density $\rho(r,t)$ at the in-plane coordinate $r$ and time $t$.

Most importantly, $S_c = -\sum_j \int u_j(r)\rho(r,t)V_j(t)d^2rdt$ describes non-local electrostatic interaction between the fermion charge density at position $r$ and the electric potential at the $j_{th}$ SC puddle, denoted by $V_j(t)$, via dimensionless potential $u_j(r) = (a^2 d/2\pi)[(r-R_j)^2+d^2]^{-3/2}$, with $d$ being the interlayer distance and $a^2$ the SC puddle area. We note that $u_j(r)$ is positive-definite and analogous to the long-range Fröhlich electron-phonon couplings in polaron physics[40]. Via the ac Josephson relation $V_j(t) = -(\hbar/2q)\partial_t \varphi_j(t)$, where $q$ is carrier charge of the LAO/STO interface ($q < 0$ for electrons and $q > 0$ for holes), one can readily realize that the coupling action $S_c$ in fact describes the interactions between the SC phases and the graphene electrons. As depicted in Fig. 4a, the evolutions of the SC phases generate time-dependent electric potentials in graphene and thus charge density fluctuations, and, reciprocally, graphene charge density fluctuations could stimulate Cooper-pair tunnelings in the JJ array and thus SC phases variations. This type of interactions is intrinsically quantum and electrodynamical owning to the Josephson effect: It is absent if the system is composed of charge-neutral quasiparticles, in thermal equilibrium, or in the classical limit $\hbar \to 0$, and can be substantially enhanced for high-frequency fluctuations.

The so-induced drag effect, which we call the Josephson-Coulomb (JC) drag effect, manifests in the coupling action. The change density fluctuations in graphene induce an effective bias current $I_j(t)$ flowing out of the $j_{th}$ SC puddle; The electric potentials in JJ array induce electric potentials $V(r,t)$ in graphene. In the semiclassical limit one has $I_j(t) = \int j(r,t)\cdot\nabla u_j(r) d^2r$ and $V(r,t) = \sum_j u_j(r)V_j(t)$, where $j(r,t)$ is the 2D charge current density in graphene. Furthermore, in the globally SC region of the JJ array, the interlayer coupling can be renormalized by the gapless Bogoliubov-Anderson-Goldstone mode[41-43] via the vertex corrections, i.e., $u_j(r) \to \gamma u_j(r)$. We have obtained the enhancement factor $\gamma(\rho,T) \approx E_J(T)/(\hbar v_F)^2 \chi_0(\rho,T)$, where $\chi_0(\rho,T)$ is the static and uniform Lindhard function of graphene electrons (Supplementary Information section 4).

In Fig. 4b, we show a schematic diagram of the JC drag processes between a one-dimensional (1D) JJ array and a graphene strip. along x-direction in the region $0 < x < L$. A uniform active current $I_G$ in a graphene strip induces the bias currents $I_j = I_G [u_j(L)-u_j(0)]$, for which the sign depends on that of the effective potential $u_j(x)$ as well as the location of the puddle (see details in Supplementary Information section 4). As the result, at each junction the induced current is in parallel to the active one, obtained by the Kirchoff's law, so that the rectification voltage develops and leads to a positive PAR. Further assuming $N$ identical JJ arrays in parallel, we estimate the PAR as $r(\rho,T) \approx aNE_J(T)/2\pi d(\hbar v_F)^2 \chi_0(\rho,T)$, so that it can be substantially enhanced by increasing $N$. In the degenerate (non-degenerate) region of graphene $T \ll |E_F|$ ($T \gg |E_F|$) [44], where $E_F$ is the Fermi energy with respect to the charge neutrality, the order of magnitude of the PAR is given by $r \sim aNE_J(T)/d\max(|E_F|,T)$. In practice, the SC puddles form a complex 2D network rather than independent 1D arrays. Nevertheless, this only influences the quantitative estimation of the channel number $N$.

According to the PAR, the JC drag exhibits unique properties distinct from the conventional Coulomb drag phenomena, especially, those based on the momentum transfer mechanism[3]. First, the sign of the PAR is positive-definite, determined by that of the effective coupling $u_j(r)$ and independent of the carrier types of both layers. Second, at constant temperature, the PAR is maximized as the graphene layer approaching the charge neutrality since the Coulomb interaction is less screened. These conclusions coincide well with our experimental result as demonstrated in Extended Data Fig. 11, where the drag resistance is almost particle-hole symmetric as the graphene layer tuned across the Dirac point. The maximum value of the drag resistance around the Dirac point should be limited by the charge puddles in graphene, attributed to the density dependence of the dielectricity of graphene electrons.

Most strikingly, for a fixed graphene carrier density, the PAR reaches maximum at zero temperature instead of vanishing since the Josephson energy $E_J(T)$ gets increased with decreasing temperature. In addition, the amplitude of the PAR is limited only by the experimental parameters and can diverge as $N/\max(|E_F|,T)$ when $\max(|E_F|,T) \to 0$ or $N \to \infty$. In our experiment, the graphene layer is deep in the degenerate region for $T < T_P$, and the temperature dependence of the PAR should arise from the Josephson energy, and, thus, $r \sim r_0[1-(T/T_P)^2]\exp(-T/T_0)$, where $r_0 \sim aNE_J(0)/d|E_F|$. Excellent consistency between the experiment and the theory is shown in Figs. 4c and 4d for various gate voltages (see Extended Data Fig. 12 for the results of another device). In contrast, neglecting the exponential factor, we obtain regression curves that substantially deviate from the experimental data (Fig. 4c). This also evidences the JJ-array superconductivity of the LAO/STO interface along with the two-step transition shown in Figs. 1d,e.

The PAR which is not attainable below $T_C$ now can be extrapolated down to zero temperature. As shown in Fig. 4e, the extracted $r_0$ exhibits non-monotonic back-gate dependence and it is as large as $10^5$ for the optimal gate value. For closed passive circuits (Extended Data Fig. 10), this means that applying an active current in graphene can induce an astonishing passive current $10^5$ times larger in the superconductor layer in close proximity. This giant amplification is reasonably due to the innumerable SC puddles ($N \gg 1$) and relatively large puddles compared to the interlayer distance ($a/d \gg 1$). In addition, $r_0$ is substantially suppressed as the LAO/STO interface is tuned away from the optimal gating, which can be possibly attributed to the gate dependence of the superfluid density[10].

The newly proposed JC drag mechanism reveals the unique role of the Josephson effect, which is a universal macroscopic quantum phenomenon in non-uniform superconductors, in shaping the interlayer multi-particle interactions. This JC drag should belong to the broad spectrum of the energy drag[45], since, at particle-hole symmetric point or at zero temperature, the drag effect is maximized instead of vanishing, in contrast to the conventional momentum drag[3]. The JC drag effect is inherently nonequilibrium because non-vanishing effective interlayer coupling requires quantum fluctuations of the SC phases in the superconductor. Such an effect should be much weaker for uniform superconductors due to substantially suppressed quantum fluctuations of the SC phases. The quantum-fluctuation-induced static interactions like the Casimir effect are commonly attributed to zero-point motions of the systems in equilibrium and manifested as thermodynamical variables[46]. More strikingly, the interaction between the SC phases in the superconductor and the electrons in the normal conductor that we discovered here serves as a prototype of the quantum-fluctuation-induced dynamical forces that can be detected only as the system is driven out of equilibrium.

From engineering perspective, our JC drag device works as a current (voltage) transformer when graphene (LAO/STO interface) is active. Moreover, it may provide an alternative way to synchronize the terahertz radiators based on large JJ arrays[12,13], since the bias current distribution in the JJ arrays can be directly controlled by that in a closely proximitized graphene layer[47]. Our findings are thus anticipated to promote further investigation of hybrid interlayer coupling via adopting other 2D systems possessing various quantum phases[48,49]. The inherent quantum fluctuations tend to give rising to novel many-body effects, which may further find applications in highly integrated modern electronics.

## References

1      Pogrebinskii, M. B. Mutual drag of carriers in a semiconductor-insulator-semiconductor system. *Sov. Phys. Semicond.* **11**, 372-376 (1977).

2   Price, P. J. Hot electron effects in heterolayers. *Physica B+C*, **117**, 750-752 (1983).
3   Narozhny, B. & Levchenko, A. Coulomb drag. *Rev. Mod. Phys.* **88**, 025003 (2016).
4   Gramila, T. J., Eisenstein, J. P., MacDonald, A. H., Pfeiffer, L. N., & West, K. W. Mutual friction between parallel two-dimensional electron systems. *Phys. Rev. Lett.* **66**, 1216 (1991).
5   Eisenstein, J. P. New transport phenomena in coupled quantum wells, *Superlattices Microstruct.* **12**, 107-114 (1992).
6   Price, A., Savchenko, A., Narozhny, B., Allison, G. & Ritchie, D. Giant fluctuations of Coulomb drag in a bilayer system. *Science* **316**, 99-102 (2007).
7   Yamamoto, M., Stopa, M., Tokura, Y., Hirayama, Y. & Tarucha, S. Negative Coulomb drag in a one-dimensional wire. *Science* **313**, 204-207 (2006).
8   Laroche, D., Gervais, G., Lilly, M. & Reno, J. 1D-1D Coulomb drag signature of a Luttinger liquid. *Science* **343**, 631-634 (2014).
9   Chen, Z. et al. Carrier density and disorder tuned superconductor-metal transition in a two-dimensional electron system. *Nat. Commun.* **9**, 4008 (2018).
10  Manca, N. et al. Bimodal phase diagram of the superfluid density in $LaAlO_3/SrTiO_3$ revealed by an interfacial waveguide resonator. *Phys. Rev. Lett.* **122**, 036801 (2019).
11  Hurand, S. et al. Josephson-like dynamics of the superconducting $LaAlO_3/SrTiO_3$ interface. *Phys. Rev. B* **99**, 104515 (2019).
12  Han, S. Bi, B. Zhang, W. & Lukens, J. E. Demonstration of Josephson effect submillimeter wave sources with increased power. *Appl. Phys. Lett.* **64**, 1424 (1994).
13  Barbara, P., Cawthorne, A. B., Shitov, S. V., & Lobb, C. J. Stimulated Emission and Amplification in Josephson Junction Arrays. *Phys. Rev. Lett.* **82**, 1963 (1999).
14  Eisenstein, J. P. & MacDonald, A. H. Bose-Einstein condensation of excitons in bilayer electron systems. *Nature* **432**, 691-694 (2004).
15  Su, J.-J. & MacDonald, A. H. How to make a bilayer exciton condensate flow. *Nat. Phys.* **4**, 799-802 (2008).
16  Nandi, D., Finck, A. D. K., Eisenstein, J. P., Pfeiffer, L. N. & West, K. W. Exciton condensation and perfect Coulomb drag. *Nature* **488**, 481-484 (2012).
17  Liu, X., Watanabe, K., Taniguchi, T., Halperin, B. I. & Kim, P. Quantum Hall drag of exciton condensate in graphene. *Nat. Phys.* **13**, 746-750 (2017).
18  Li, J., Taniguchi, T., Watanabe, K., Hone, J. & Dean, C. Excitonic superfluid phase in double bilayer graphene. *Nat. Phys.* **13**, 751-755 (2017).
19  Du, L. et al. Coulomb drag in topological wires separated by an air gap. *Nat. Electron.* **4**, 573-578 (2021).
20  Giaever, I. Magnetic coupling between two adjacent type-II superconductors. *Phys. Rev. Lett.* **15**, 825 (1965).
21  Giaever, I. Flux pinning and flux-flow resistivity in magnetically coupled superconducting films. *Phys. Rev. Lett.* **16**, 460 (1966).
22  Averin, D. V., Korotkov, A. N. & Nazarov, Y. V. Transport of electron-hole pairs in arrays of small tunnel junctions. *Phys. Rev. Lett.* **66**, 2818 (1991).
23  Duan, J.-M. & Yip, S. Supercurrent drag via the Coulomb interaction. *Phys. Rev. Lett.* **70**, 3647-3650 (1993).
24  Giordano, N. & Monnier, J. D. Cross-talk effects in superconductor-insulator-normal-metal trilayers. *Phys. Rev. B* **50**, 9363-9368 (1994).
25  Huang, X., Bazàn, G. & Bernstein, G. H. Observation of supercurrent drag between normal metal and superconducting films. *Phys. Rev. Lett.* **74**, 4051-4054 (1995).
26  Shimshoni, E. Role of vortices in the mutual coupling of superconducting and normal-metal films. *Phys. Rev. B* **51**, 9415-9418 (1995).

27. Shimada, H. & Delsing, P. Current mirror effect and correlated Cooper-pair transport in coupled arrays of small Josephson junctions. *Phys. Rev. Lett.* **85**, 3253 (2000).
28. Zou, Y., Refael, G. & Yoon, J. Investigating the superconductor-insulator transition in thin films using drag resistance: theoretical analysis of a proposed experiment. *Phys. Rev. B* **80**, 180503 (2009).
29. Levchenko, A. & Norman, M. R. Proposed Giaever transformer to probe the pseudogap phase of cuprates. *Phys. Rev. B* **83**, 100506 (2011).
30. Andreev, A. & Bashkin, E. Three-velocity hydrodynamics of superfluid solutions. *Sov. Phys. JETP* **42**, 164-167 (1976).
31. Alpar, M., Langer, S. A. & Sauls, J. Rapid postglitch spin-up of the superfluid core in pulsars. *Astrophys. J.* **282**, 533-541 (1984).
32. Kim, S. *et al.* Coulomb drag of massless fermions in graphene. *Phys. Rev. B* **83**, 161401 (2011).
33. Gorbachev, R. V. *et al.* Strong Coulomb drag and broken symmetry in double-layer graphene. *Nat. Phys.* **8**, 896-901 (2012).
34. Gamucci, A. *et al.* Anomalous low-temperature Coulomb drag in graphene-GaAs heterostructures. *Nat. Commun.* **5**, 5824 (2014).
35. Reyren, N. *et al.* Superconducting interfaces between insulating oxides. *Science* **317**, 1196-1199 (2007).
36. Caviglia, A. *et al.* Electric field control of the $LaAlO_3/SrTiO_3$ interface ground state. *Nature* **456**, 624-627 (2008).
37. Bert, J. A. *et al.* Direct imaging of the coexistence of ferromagnetism and superconductivity at the $LaAlO_3/SrTiO_3$ interface. *Nat. Phys.* **7**, 767-771 (2011).
38. Prozorov, R. & Giannetta, R. W. Magnetic penetration depth in unconventional superconductors. *Supercon. Sci. Technol.* **19**, R41 (2006).
39. Kapitulnik, A., Kivelson, S. A. & Spivak, B. Colloquium: anomalous metals: Failed superconductors. *Rev. Mod. Phys.* **91**, 011002 (2019).
40. Alexandrov, A. S. & Kornilovitch, P. E. Mobile small polaron. *Phys. Rev. Lett.* **82**, 807 (1999).
41. Bogoljubov, N. N., Tolmachov, V. V. & Širkov, D. V. A new method in the theory of superconductivity. *Fortschr. Phys.* **6**, 605-682 (1958).
42. Anderson, P. W. Coherent excited states in the theory of superconductivity: Gauge invariance and the Meissner effect. *Phys. Rev.* **110**, 827 (1958).
43. Goldstone, J. Field theories with Superconductor solutions. *Nuovo Cim.* **19**, 154-164 (1961).
44. Ramezanali, M. R., Vazifeh, M. M., Asgari, R., Polini, M., & MacDonald, A. H. Finite-temperature screening and the specific heat of doped graphene sheets. *J. Phys. A: Math. Theor.* **42**, 214015 (2009).
45. Berdanier, W., Scaffidi, T. & Moore J. E. Energy drag in particle-hole symmetric systems as quantum quench. *Phys. Rev. Lett.* **123**, 246603 (2019).
46. Casimir, H. B. G. On the attraction between two perfectly conducting plates. *Proc. K. Ned. Akad. Wet. B* **51**, 793-795 (1948).
47. Galin, M. A. *et al.* Direct visualization of phase-locking of large Josephson junction arrays by surface electromagnetic waves. *Phys. Rev. Appl.* **14**, 024051 (2020).
48. Saito, Y., Nojima, T. & Iwasa, Y. Highly crystalline 2D superconductors. *Nat. Rev. Mater.* **2**, 16094 (2017).
49. Gong, C. & Zhang, X. Two-dimensional magnetic crystals and emergent heterostructure devices. *Science* **363**, 6428 (2019).

## Methods

**Device fabrication.** LAO/STO heterostructures were fabricated using pulsed laser deposition following those described in our previous study[50]. The ultra-flat surface and the good interfacial conductivity were verified prior to the following procedures. Monolayer graphene sheets were grown on Cu foils by chemical vapor deposition[51], and then transferred directly onto the LAO/STO surface as previously reported[52]. For the transport measurements, Al (Au) wires were connected to the LAO/STO interface (the graphene layer) using ultrasonic welding (silver conductive paint). Ti/Au contacts with a thickness of 5/50 nm were fabricated on the back side of the STO substrates to apply the back-gate voltage.

**Electronic transport measurements.** The electronic transport measurements were performed in an Oxford Instruments Triton Dilution Refrigerator. During the drag measurements, direct current (DC) mode was adopted to avoid possible influence of capacitive reactance in alternating current (AC) measurements[33]. Keithley 6220/6221 and 2182A were employed to supply the currents in the drive layer and measure the voltage drops in the drag layer, respectively. To eliminate the voltage background of 2182A, the current was applied using a bipolar mode, and the voltage was obtained by taking the average of the measured voltages at positive and negative currents.

## References

50   Liang, H. *et al.* Nonmonotonically tunable Rashba spin-orbit coupling by multiple-band filling control in SrTiO$_3$-based interfacial d-electron gases. *Phys. Rev. B* **92**, 075309 (2015).
51   Li, X. *et al.* Large-area synthesis of high-quality and uniform graphene films on copper foils. *Science* **324**, 1312-1314 (2009).
52   Cheng, L. *et al.* Photoconductivity of graphene in proximity to LaAlO$_3$/SrTiO$_3$ heterostructures: phenomenon and photosensor applications. *Phys. Rev. Appl.* **6**, 014005 (2016).

## Acknowledgments

We thank Dr. Fengcheng Wu for helpful discussions. This work was supported by the National Natural Science Foundation of China (Grant Nos. 92165201, 11804326, 11974324, U1832151, 12074039), the Strategic Priority Research Program of Chinese Academy of Sciences (Grant No. XDC07010000), the Anhui Initiative in Quantum Information Technologies (Grant No. AHY170000), Hefei Science Center CAS (Grant No. 2020HSC-UE014), and the Fundamental Research Funds for the Central Universities (Grant No. WK3510000013). Part of this work was carried out at the USTC Center for Micro and Nanoscale Research and Fabrication.

**Author contributions**

C.Z. and L.L. designed and supervised the work; R.T. and L.L. performed the experiments with assistance from X.F., L.G., L.Z. and Y.Y.; H.-Y X. conceived the theoretical model. L.L., R.T., H.-Y.X. and C.Z. analyzed the data and wrote the manuscript; Z.Z. contributed to data interpretation and presentation. All authors contributed to the scientific discussion and manuscript revisions.

**Competing financial interests**

The authors declare no competing financial interests.

**Figures**

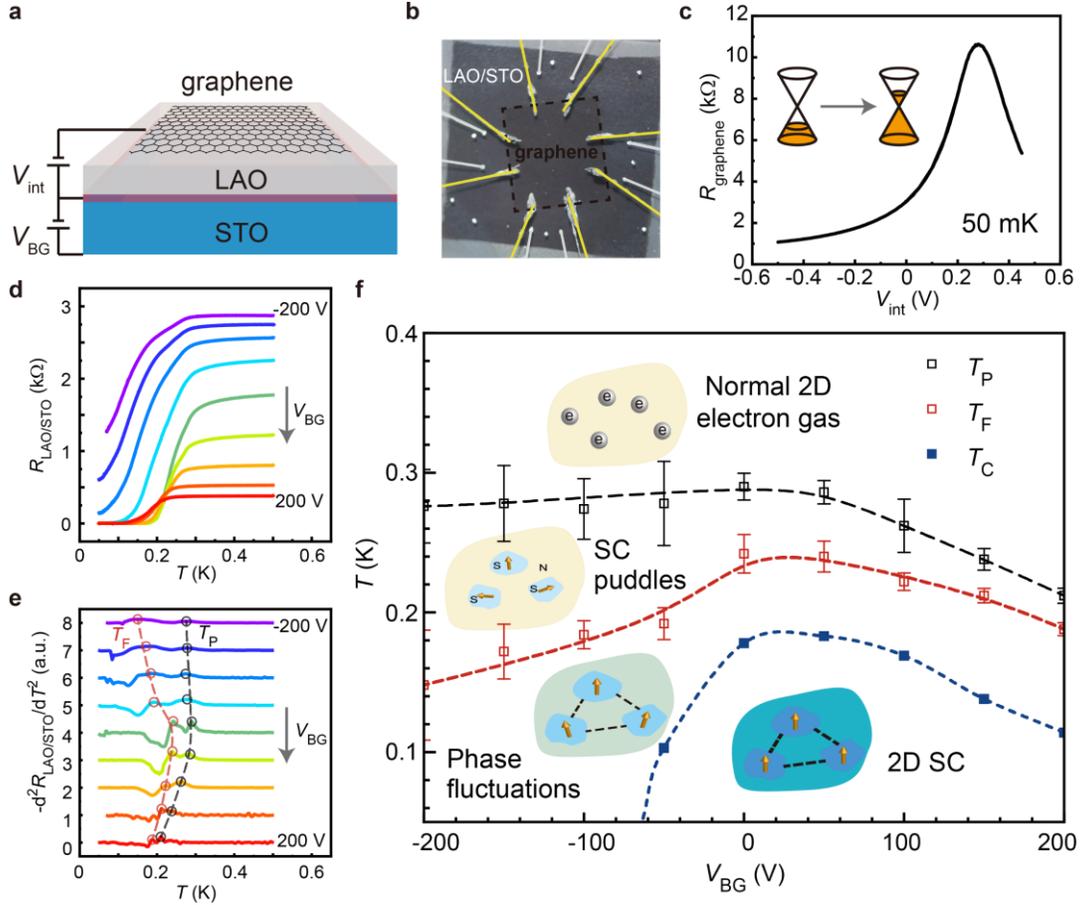

**Fig. 1 | Device and intralayer transport characterizations. a**, Schematic illustration and **b**, optical microscope image of the G-LAO/STO device. **c**, Resistance of graphene ($R_{graphene}$) as a function of interlayer-gate voltage ($V_{int}$) measured at 50 mK. **d**, Resistance of the LAO/STO interface ($R_{LAO/STO}$) as a function of temperature ($T$) for back-gate voltage ($V_{BG}$) varying from -200 to 200 V in increments of 50 V. **e**, Second derivative of the $R$-$T$ curves ($-d^2R_{LAO/STO}/dT^2$) shown in **d**. For a fixed $V_{BG}$, the two maxima are denoted as $T_F$ (red circles) and $T_P$ (black circles). The dashed lines indicate the traces of $T_F$ and $T_P$ as $V_{BG}$ varies. **f**, Low-temperature $V_{BG}$-$T$ phase diagram of the LAO/STO interface obtained from **d** and **e**. We define $T_C$ (the blue squares) the temperature at which the resistance drops to 1% of the normal state resistance at $T$ = 400 mK. $T_F$ (the black hollow squares) and $T_P$ (the red hollow squares) are the second-derivative maxima in **e** and the error bars are evaluated by the corresponding peak widths. The dashed lines estimate phase boundaries and the schematics of the phases are depicted.

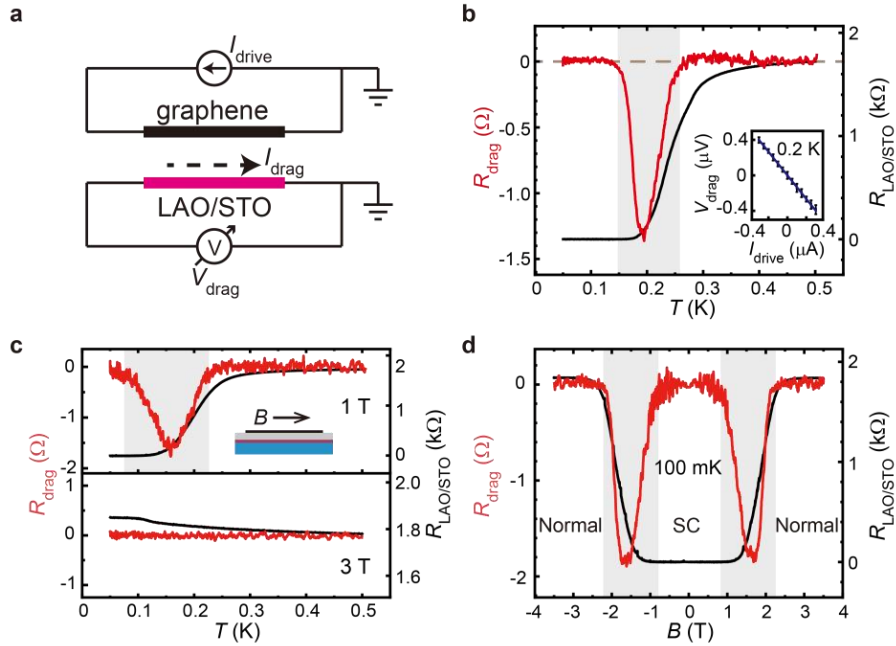

**Fig. 2 | Interlayer drag resistances. a**, Schematic illustration of the set-up for drag measurements. Active current $I_{drive}$ is applied to the graphene layer, passive voltage drop $V_{drag}$ is measured at the LAO/STO interface, and the passive current is defined by $I_{drag} = -V_{drag}/R_{LAO/STO}$. **b-d** show the results for zero $V_{BG}$ (see Fig. 1a). **b**, Drag resistance ($R_{drag} = V_{drag}/I_{drive}$) and $R_{LAO/STO}$ as functions of $T$ in the absence of magnetic field. The inset shows perfect linear dependence of $V_{drag}$ upon $I_{drive}$ for $|I_{drive}| < 0.4$ μA. **c**, $R_{drag}$ and $R_{LAO/STO}$ as functions of $T$ for in-plane magnetic field $B = 1$ T (upper panel) and $B = 3$ T (lower panel). **d**, $R_{drag}$ and $R_{LAO/STO}$ as functions of in-plane $B$ at $T = 100$ mK.

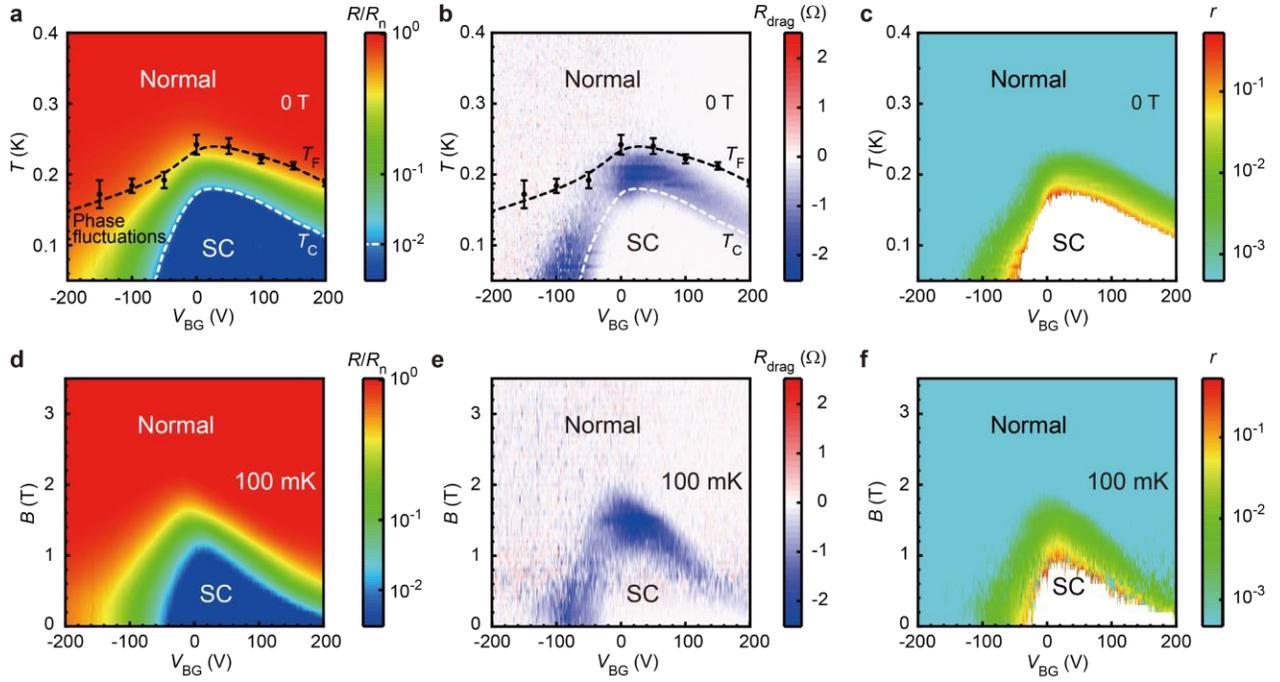

**Fig. 3 | Phase diagrams of the interlayer drag effect. a-c**, Normalized resistance of the LAO/STO interface ($R_{LAO/STO}/R_n$), $R_{drag}$, and dimensionless drag coefficient of PAR ($r = -R_{drag}/R_{LAO/STO}$) as functions of $V_{BG}$ and $T$ in the absence of magnetic field, respectively. $R_n$ is the normal state resistance of the LAO/STO interface at $T = 400$ mK. In **a** and **b**, $T_C$ (white dashed lines) and $T_F$ (black squares and dashed lines) and the error bars are defined in Fig. 1f. **d-f**, $R_{LAO/STO}/R_n$, $R_{drag}$, and $r$ as functions of $V_{BG}$ and in-plane $B$ at $T = 100$ mK, respectively.

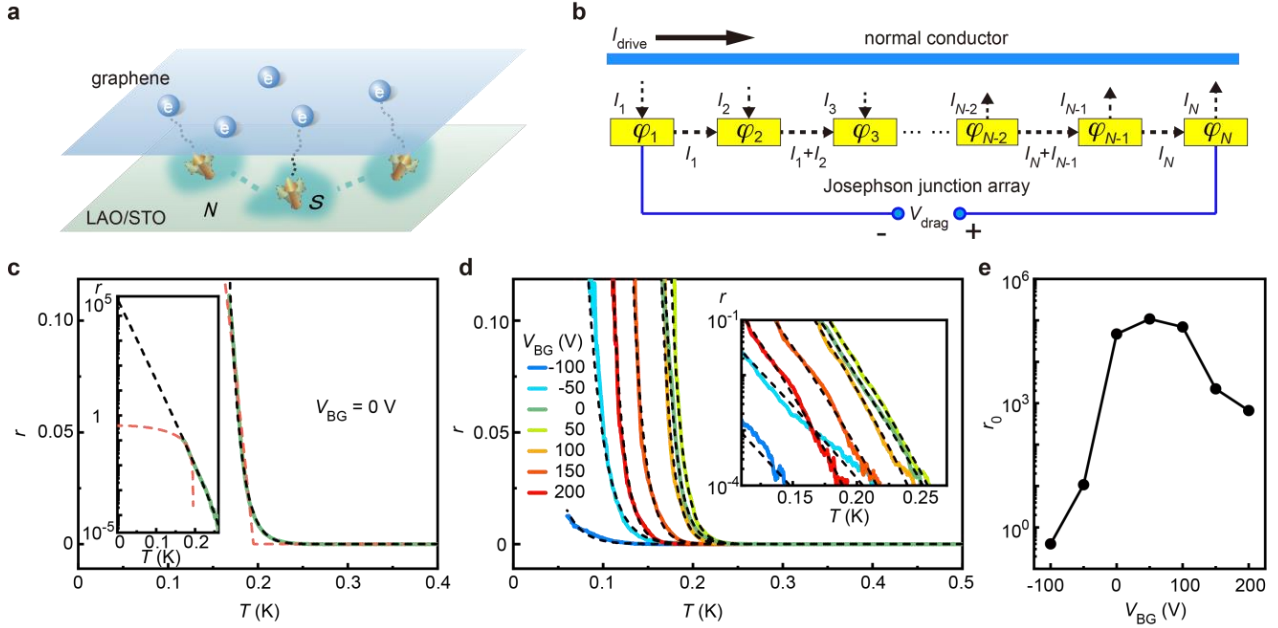

**Fig. 4 | Josephson-Coulomb (JC) drag effect below $T_P$. a**, Schematic illustration of the long-range couplings between the phases of the order parameters of the SC puddles in LAO/STO interface and the graphene electrons mediated by Coulomb interactions below $T_P$. Charge-density temporal fluctuations in graphene accelerate the phase velocity of a SC puddle in LAO/STO interface, and, conversely, phase velocity of a SC puddle generates an electric potential acting on graphene electrons. **b**, Schematics of JC drag between a JJ array and a normal conductor in 1D geometry. Active current $I_G$ is applied in the normal conductor and an induced current $I_j(t)$ flows in or out of the $j$th puddle depending on the location. At each junction, the passive bias current is in parallel to the active and the rectification voltage leads to a positive PAR. **c**, Drag coefficient of PAR ($r$) as a function of $T$ for $V_{BG} = 0$ V and $B = 0$ T. We use the functions $r(T) = r_0[1-(T/T^*)^2]$ (orange dashed line) and $r(T) = r_0[1-(T/T_P)^2]\exp(-T/T_0)$ (black dashed line) to fit the data. The inset shows the semi-log plots in the low-temperature region. **d**, $r$ as a function of $T$ for various $V_{BG}$. We fit the data using the function $r(T) = r_0[1-(T/T_P)^2]\exp(-T/T_0)$. The inset shows the semi-log plots. **e**, Zero-temperature drag coefficient $r_0$ as a function of $V_{BG}$ extracted from the fitting results in **d**.